\documentstyle[12pt,epsfig]{article}
\title{ The Reach of LHC (CMS) for Models with Effective 
Supersymmetry and Nonuniversal Gaugino Masses}
\author{\large S.I.~Bityukov$^1$ and  N.V.~Krasnikov  \\[3mm]
\em Institute for Nuclear Research RAS, \\
\em Moscow, 117312, Russia  }
\date{}
\begin{document}
\maketitle
\begin{abstract}
We investigate squark and gluino  pair production
at LHC (CMS) with subsequent decays into quarks, leptons and LSP 
in models with effective 
supersymmetry where third generation of squarks is relatively light
while the first two generations of squarks are heavy.
We consider the general case of nonuniversal gaugino masses.
Visibility of signal by an excess over SM background 
in $(n \geq 2)jets + (m \geq 0)leptons + E^{miss}_T$ events
depends rather strongly on the relation between LSP, second
neutralino, gluino and 
squark masses and it decreases with the increase of LSP mass. 
\end{abstract}
\vspace{1cm}
\bigskip

\noindent
\rule{3cm}{0.5pt}\\
$^1$~~Institute for High Energy Physics, Protvino, Russia

\section{Introduction}

One of the LHC supergoals is the discovery of the supersymmetry.
In particular, it is very important to investigate a possibility
to discover strongly interacting superparticles (squarks and gluino). 
In ref.\cite{1} (see, also references \cite{2}) the LHC squark 
and gluino  
discovery potential
has been investigated within the minimal SUGRA-MSSM framework
\cite{3} where all 
sparticle masses are determined mainly by two parameters: $m_0$ (common
squark and slepton mass at GUT scale) and $m_{1 \over 2}$ (common
gaugino mass at GUT scale). 
The signature used for the search for squarks and gluino  at LHC is 
$(n \ge 2)jets$ + $(m \ge 0)$ leptons + $E^{miss}_T$ events. 
The conclusion of ref.~\cite{1}  is that LHC is able 
to detect squarks and gluino  with masses up to 
(2 - 2.5) TeV.  
In ref.~\cite{4} the LHC SUSY discovery potential has been investigated
for the case of nonuniversal gaugino masses with universal squark masses
for the first, second and third generations. The conclusion of the
ref.~\cite{4} is that visibility of signal by an excess over SM
background in $(n \geq 2)jets + E^{miss}_T$ events depends rather strongly
on the relation between LSP, gluino and squark masses and it decreases
with the increase of LSP mass. For relatively heavy LSP mass closed to
squark or gluino masses and for 
$(m_{\tilde q}, m_{\tilde g}) \geq 1.5~TeV$ signal is too small
to be observable.

In this paper we investigate the squark and gluino pair production 
at LHC (CMS) with subsequent decays into quarks, leptons and LSP in models
with effective supersymmetry~\cite{5} where third generation of squarks
is relatively light while the first two generations of squarks are heavy.
Note that in ref.~\cite{15} ATLAS detector discovery potential of
mSUGRA with focus point effective supersymmetry for $tan \beta = 10$
and $\mu < 0$ has been studied for signature $n \ge 2$ jets $+~1$
isolated lepton plus $E^{miss}_T$. 
We consider the general case of nonuniversal gaugino masses. Visibility
of signal by an excess over SM background  in
$(n \geq 2)$jets + $(m \geq 0)$ leptons + $E^{miss}_T$ events
depends rather strongly on the relation
between LSP, second neutralino, gluino and squark masses and it decreases
with the increase of LSP mass. 

Despite the simplicity of the  SUGRA-MSSM framework it is a very particular
model. The mass formulae for sparticles in  SUGRA-MSSM model are derived
under the assumption that at GUT scale ($M_{GUT} \approx 2 \cdot 10^{16}$~GeV) 
soft supersymmetry breaking terms are universal. However, in general,
we can expect that real sparticle masses can differ in a drastic way 
from sparticle masses pattern of SUGRA-MSSM model due to many reasons,
see for instance refs.~\cite{6,7,8,9,10}.
Therefore, it is more appropriate to investigate the LHC SUSY discovery 
potential in a model-independent way. 

The cross section 
for the production of strongly interacting superparticles
\begin{equation}
pp \rightarrow \tilde{g}\tilde{g}, \tilde{q}\tilde{g}, \tilde{q} \tilde{q}
\end{equation}
depends on gluino and squark masses.
Within SUGRA-MSSM  model the following approximate relations among sparticle 
masses take place:
\begin{equation}
m^2_{\tilde{q}} \approx m^2_0 + 6 m^2_{1 \over 2},
\end{equation}
\begin{equation}
m_{\tilde{\chi}^0_1} \approx 0.45 m_{1 \over 2},
\end{equation}
\begin{equation}
m_{\tilde{\chi}^0_2} \approx m_{\tilde{\chi}^{\pm}_1} \approx 
2m_{\tilde{\chi}^0_1},
\end{equation}
\begin{equation}
m_{\tilde{g}} \approx 2.5m_{1 \over 2}
\end{equation}
The decays of squarks and gluino depend on the relation among squark and 
gluino masses. For $m_{\tilde{q}} > m_{\tilde{g}}$ squarks decay mainly 
into gluino and quarks

\begin{itemize}

\item
$\tilde{q} \rightarrow \tilde{g} q$

\end{itemize}  
and gluino decays mainly into quark-antiquark pair and gaugino

\begin{itemize}

\item
$\tilde{g} \rightarrow q \bar{q} \tilde{\chi}^0_i$

\end{itemize}
\begin{itemize}

\item
$\tilde{g} \rightarrow q \bar{q}^{'} \tilde{\chi}^{\pm}_{1}$

\end{itemize}

For $m_{\tilde{q}} < m_{\tilde{g}}$ gluino decays mainly into squarks and 
quarks 

\begin{itemize}

\item

$\tilde{g} \rightarrow \bar{q} \tilde{q}, q \bar{\tilde{q}}$

\end{itemize}

whereas squarks decay mainly into quarks and gaugino

\begin{itemize}

\item
$ \tilde{q} \rightarrow   q \tilde{\chi^0_{i}}$

\end{itemize}
\begin{itemize}
\item
$\tilde{q} \rightarrow q^{'} \tilde{\chi}^{\pm}_1$

\end{itemize} 

The lightest chargino $\tilde \chi^{\pm}_1$ has several leptonic decay modes
giving a lepton and missing energy:

three-body decay

\begin{itemize}

\item 
$\tilde \chi^{\pm}_1 \longrightarrow \tilde \chi^0_1  + l^{\pm} + \nu$,

\end{itemize}

two-body decays

\begin{itemize}

\item
$\tilde \chi^{\pm}_1  \longrightarrow  \tilde l^{\pm}_{L,R} + \nu$,

\hspace{16mm}  $\hookrightarrow \tilde \chi^0_1 + l^{\pm}$

\item
$\tilde \chi^{\pm}_1 \longrightarrow \tilde \nu_L + l^{\pm}$,

\hspace{16mm} $ \hookrightarrow \tilde \chi^0_1 + \nu$

\item
$\tilde \chi^{\pm}_1 \longrightarrow \tilde \chi^0_1 + W^{\pm}$.

\hspace{26mm} $ \hookrightarrow l^{\pm} + \nu$

\end{itemize}

Leptonic decays of $\tilde \chi^0_2$ give two leptons and missing 
energy:

three-body decays

\begin{itemize}
\item 
$\tilde \chi^0_2 \longrightarrow \tilde \chi^0_1 + l^+ l^-$,

\item
$\tilde \chi^0_2 \longrightarrow \tilde \chi^{\pm}_1 + l^{\mp} + \nu$,

\hspace{16mm} $ \hookrightarrow \tilde \chi^0_1 + l^{\pm} + \nu$

\end{itemize}

two-body decay

\begin{itemize}
\item
$\tilde \chi^0_2 \longrightarrow \tilde l^{\pm}_{L,R} + l^{\mp}$.

\hspace{16mm} $ \hookrightarrow \tilde \chi^0_1 + l^{\pm}$

\end{itemize}

As a result of chargino and second neutralino leptonic decays besides 
classical signature

\begin{itemize}
\item
$(n \geq 2)$ jets plus $E^{miss}_{T}$
\end{itemize}
signatures

\begin{itemize}
\item
$(n \geq 2)$ jets plus $(m \geq 1)$ leptons plus $E^{miss}_{T}$
\end{itemize}
with leptons and jets in final state arise.
As mentioned above, these signatures have been used in ref.~\cite{1}
for investigation of LHC(CMS) squark and gluino discovery 
potential within SUGRA-MSSM  model, in which  gaugino masses 
$m_{\tilde \chi^0_1}$, $m_{\tilde \chi^0_2}$ are
determined mainly by a common gaugino mass $m_{1 \over 2}$. 

In our  
study we consider the models with effective supersymmetry~\cite{5}
in which first and second squark generations are heavy while 
the third squark generation is relatively light. Models with
effective supersymmetry solve in natural way the problems with 
flavour-changing neutral currents, lepton flavor violation,
electric dipole moments of electron and neutron and proton decay.
In such models there are two mass scales: gauginos, higgsinos
and third generation squarks are rather light to stabilize
electroweak scale, while the first two generations of squarks
and sleptons are heavy with masses $\approx(5-20)~TeV$. 
We investigate the general case when the relation among
gaugino masses is arbitrary. We study 
the detection of supersymmetry using classical signature 
$(n \geq 2)$ jets + $(m \geq 0)$leptons + $E^{miss}_{T}$. 
We find that the SUSY discovery potential depends rather 
strongly on the relation among squarks, gluino, LSP and second
neutralino masses and it decreases with the increase of LSP mass.
 
\section{Simulation of detector response}

Our simulations are made at the particle level with parametrized
detector responses based on a detailed detector simulation. 
To be concrete our estimates have been made for the CMS(Compact Muon 
Solenoid)  detector. The CMS detector simulation 
program CMSJET~7.4~\cite{11} is used.
The main aspects of the CMSJET relevant to our study are the 
following.

\begin{itemize}
\item
Charged particles are tracked in a 4 T magnetic field. 90 percent 
reconstruction  efficiency per charged track with $p_T > 1$ GeV within 
$|\eta| <2.5$ is assumed. 

\end{itemize}

\begin{itemize}
\item
The geometrical acceptances for $\mu$ and $e$ are $|\eta| <2.4$ and 2.5, 
respectively. The lepton number is smeared according to parametrizations 
obtained from full GEANT simulations. For a 10 GeV lepton the momentum 
resolution $\Delta p_T/p_T$ is better than one percent over the full $\eta$ 
coverage. For a 100 GeV lepton the resolution becomes $\sim (1 - 5) \cdot 
10^{-2}$ depending on $\eta$. We have assumed a 90 percent triggering 
plus reconstruction efficiency per lepton within the geometrical 
acceptance of the CMS detector.  

\end{itemize} 

\begin{itemize}
\item
The electromagnetic calorimeter of CMS extends up to $|\eta| = 2.61$. There 
is a pointing crack in the ECAL barrel/endcap transition region 
between $|\eta| = 1.478 - 1.566$ (6 ECAL crystals). The hadronic calorimeter 
covers $|\eta| <3$. The Very Forward calorimeter extends from $|\eta| <3$ 
to $|\eta| < 5$. Noise terms have been simulated with Gaussian distributions 
and zero suppression cuts have been applied. 
\end{itemize}

\begin{itemize}
\item
$e/\gamma$ and hadron shower development are taken into account by 
parametrization of the lateral and longitudinal profiles of showers. The 
starting point of a shower is fluctuated according to an exponential 
law.

\end{itemize}

\begin{itemize}
\item
For jet reconstruction we have used a slightly modified UA1 Jet Finding 
Algorithm, with a cone size of $\Delta R = 0.8$ and 25 GeV transverse 
energy threshold on jets. 

\end{itemize}

\section{Backgrounds. SUSY kinematics}
  
All SUSY processes with full particle spectrum, couplings,
production cross section and decays are generated with ISAJET~7.42,
ISASUSY~\cite{12}. The Standard Model backgrounds are generated 
by Pythia~5.7~\cite{13}. 

The following SM processes give the main contribution to the background:

\noindent
$WZ,~ZZ,~t \bar t,~Wtb,~Zb \bar b,~b \bar b$ and QCD $(2 \rightarrow 2)$ 
processes. 

As it has been mentioned above in this paper we consider signature
$(n \geq m)$ jets plus $(m \geq k)$ isolated leptons plus $E^{miss}_{T}$,
where $m=2,3,4$ and $k=0,1,2,3$. Namely we have considered signatures~:
\begin{itemize}
\item $(n \geq m)$ jets plus $E^{miss}_T$,
\item $(n \geq m)$ jets plus $E^{miss}_T$ plus no isolated leptons,
\item $(n \geq m)$ jets plus $E^{miss}_T$ plus $1$ isolated lepton,
\item $(n \geq m)$ jets plus $E^{miss}_T$ plus $l^+l^-$ pair of 
isolated leptons,
\item $(n \geq m)$ jets plus $E^{miss}_T$ plus $l^{\pm}l^{\pm}$ pair of
isolated leptons,
\item $(n \geq m)$ jets plus $E^{miss}_T$ plus $3$ isolated leptons.
\end {itemize}

For leptons we use cut 
$P_{lT} \equiv \sqrt{p^2_{l1} + p^2_{l2}} \geq P_{lT_0} = 20 GeV$.
Our definition of isolated lepton coincides with the definition
used in CMSJET code~\cite{11}. We use two sets of cuts (a and b) 
for $E_T^{miss}$ and $E_{Tjet,k}$ $(k=1,2,3,4)$.  Cuts a and b are presented
in tables 1 and 2, correspondingly. Besides we require that 
$\frac{N_s}{N_b} \geq 0.25$. We have calculated SM backgrounds for
different values of 
$E^0_{Tjet1},~E^0_{Tjet2},~E^0_{Tjet3},~E^0_{Tjet4},~E^0_{Tmiss}$
using PYTHIA~5.7 code~\cite{13}. We have considered two values of
$tan \beta = 5$ and $tan \beta = 35~~(tan \beta \equiv \frac{<H_t>}{<H_b>})$.
We considered both cases of heavy and relatively light gluino.
We considered different values of LSP and second neutralino masses.
In our calculations we took the value of the masses of the first and
second squark generations equal to $m_{\tilde q_{1,2}} = 3800 GeV$. However
for $m_{\tilde q_{1,2}} \geq 3000 GeV$ the result practically do not
depend on the value of $m_{\tilde q_{1,2}}$.

\section{Results}

The results of our calculations are presented in tables (3-24). 
In estimation of LHC (CMS) SUSY discovery potential we took total
luminosity equal to $L_t = 10^5pb^{-1}$. There is a crucial
difference between ``future'' experiment and ``real'' experiment~\cite{14}. 
In the ``real'' experiment the total number of events $N_{ev}$ is 
a given number and we compare it with expected $N_b$ when we test 
the validity 
of standard physics. In the condition of the ``future'' experiment
we know only the average number of the background events $N_b$ and
the average number of signal events $N_s$, so we have to compare the
Poisson distributions $P(n,N_b)$ and $P(n,N_b+N_s)$ to determine
the probability to find new physics in future experiment. 
According to common definition new physics discovery potential
corresponds to the case when the probability that background can imitate
signal is less than $5\cdot\sigma$ or in terms of the probability
less than $\Delta = 5.6\cdot10^{-7}$. So we require that the probability
$\beta(\Delta)$ of the background fluctuations for $n > n_0(\Delta)$ 
is less than $\Delta$, namely, 

$\beta(\Delta) =  
\displaystyle \sum ^{\infty}_{n=n_0({\Delta})+1} P(n,N_b) \leq \Delta$

The discovery probability $1 - \alpha(\Delta)$
that the number of signal events will be bigger than
$n_0(\Delta)$ is equal to

$1 - \alpha(\Delta) = 
\displaystyle \sum^{\infty}_{n = n_0(\Delta) + 1}P(n,N_b+N_s)$.

We require that $1 - \alpha(\Delta) \geq 0.5$.

As it follows from our results for fixed 
values of squark and gluino masses the visibility of signal decreases with 
the increase of the LSP mass. This fact has trivial explanation. Indeed, in the 
rest frame of squark or gluino the jets spectrum becomes more soft with the 
increase of LSP mass. Besides in parton model pair produced squarks and 
gluino are produced with total transverse momentum closed to zero. For high 
LSP masses partial cancellation of missing transverse momenta from two 
LSP particles takes place. 

Note that for the case of relatively light $3^{rd}$ squark generation 
$b$-quarks in the final state dominate. However, in our calculations
we have not used b-tagging to suppress the background and to make signal 
more observable.

\section{Conclusion}

In this paper we have presented the results of the investigation 
of LHC (CMS) SUSY discovery potential for the models with 
effective supersymmetry. We have considered general case 
of nonuniversal gaugino masses.
We have found that the visibility of signal by an excess over SM background
in $jets + isolated~leptons + E^{miss}_T$ events depends rather 
strongly on the relation 
between LSP, second neutralino, gluino and $3^{rd}$ generation squark 
masses and it decreases with the increase of LSP mass. 
For the case when gluino is heavy it would be possible to detect SUSY for
$3^{rd}$ generation squark masses up to $1~TeV$.

\begin{center}
 {\large \bf Acknowledgments}
\end{center}

\par
We are  indebted to 
the participants of Daniel Denegri seninar on physics 
simulations at LHC for useful comments. 
This work has been supported by RFFI grant 99-02-16956 and
INTAS-CERN 377.


\begin{table}[h]
\small
    \caption{Cuts a. }
    \label{tab.1}
\begin{center}
\begin{tabular}{|r|r|r|r|r|r|}
\hline
\# of cut&$p_{t1}$ [GeV]&$p_{t2}$ [GeV]&$p_{t3}$ [GeV]&$p_{t4}$ [GeV]& 
$E_t^{miss}$ [GeV] \\ 
\hline
    1  &   40.0 &    40.0 &    40.0 &    40.0  &  200.0 \\
    2  &  100.0 &   100.0 &   100.0 &   100.0  &  200.0 \\
    3  &  100.0 &   150.0 &   150.0 &   150.0  &  200.0 \\
    4  &   50.0 &   100.0 &   100.0 &   100.0  &  200.0 \\
    5  &  200.0 &   200.0 &   200.0 &   200.0  &  400.0 \\
    6  &  200.0 &   300.0 &   300.0 &   300.0  &  400.0 \\
    7  &  100.0 &   200.0 &   200.0 &   200.0  &  400.0 \\
    8  &  300.0 &   300.0 &   300.0 &   300.0  &  600.0 \\
    9  &  300.0 &   450.0 &   450.0 &   450.0  &  600.0 \\
   10  &  150.0 &   300.0 &   300.0 &   300.0  &  600.0 \\
   11  &  400.0 &   400.0 &   400.0 &   400.0  &  800.0 \\
   12  &  400.0 &   600.0 &   600.0 &   600.0  &  800.0 \\
   13  &  200.0 &   400.0 &   400.0 &   400.0  &  800.0 \\
   14  &  500.0 &   500.0 &   500.0 &   500.0  & 1000.0 \\
   15  &  500.0 &   750.0 &   750.0 &   750.0  & 1000.0 \\
   16  &  250.0 &   500.0 &   500.0 &   500.0  & 1000.0 \\
   17  &  600.0 &   600.0 &   600.0 &   600.0  & 1200.0 \\
   18  &  600.0 &   900.0 &   900.0 &   900.0  & 1200.0 \\
   19  &  300.0 &   600.0 &   600.0 &   600.0  & 1200.0 \\
\hline
\end{tabular}
    \end{center}
\end{table}

\begin{table}[b]
\small
    \caption{Cuts b.}
    \label{tab.2}
\begin{center}
\begin{tabular}{|r|r|r|r|r|r|}
\hline
\# of cut&$p_{t1}$ [GeV]&$p_{t2}$ [GeV]&$p_{t3}$ [GeV]&$p_{t4}$ [GeV]& 
$E_t^{miss}$ [GeV] \\ 
\hline
    1 &    40.0 &    40.0 &    40.0 &    40.0 &   200.0 \\
    2 &   100.0 &   125.0 &   150.0 &   150.0 &   200.0 \\
    3 &   166.7 &   208.3 &   250.0 &   250.0 &   200.0 \\
    4 &   233.3 &   291.7 &   350.0 &   350.0 &   200.0 \\
    5 &   300.0 &   375.0 &   450.0 &   450.0 &   200.0 \\
    6 &   100.0 &   125.0 &   150.0 &   150.0 &   400.0 \\
    7 &   166.7 &   208.3 &   250.0 &   250.0 &   400.0 \\
    8 &   233.3 &   291.7 &   350.0 &   350.0 &   400.0 \\
    9 &   300.0 &   375.0 &   450.0 &   450.0 &   400.0 \\
   10 &   100.0 &   125.0 &   150.0 &   150.0 &   600.0 \\
   11 &   166.7 &   208.3 &   250.0 &   250.0 &   600.0 \\
   12 &   233.3 &   291.7 &   350.0 &   350.0 &   600.0 \\
   13 &   300.0 &   375.0 &   450.0 &   450.0 &   600.0 \\
   14 &   100.0 &   125.0 &   150.0 &   150.0 &   800.0 \\
   15 &   166.7 &   208.3 &   250.0 &   250.0 &   800.0 \\
   16 &   233.3 &   291.7 &   350.0 &   350.0 &   800.0 \\
   17 &   300.0 &   375.0 &   450.0 &   450.0 &   800.0 \\
   18 &   100.0 &   125.0 &   150.0 &   150.0 &  1000.0 \\
   19 &   166.7 &   208.3 &   250.0 &   250.0 &  1000.0 \\
   20 &   233.3 &   291.7 &   350.0 &   350.0 &  1000.0 \\
   21 &   300.0 &   375.0 &   450.0 &   450.0 &  1000.0 \\
   22 &   100.0 &   125.0 &   150.0 &   150.0 &  1200.0 \\
   23 &   166.7 &   208.3 &   250.0 &   250.0 &  1200.0 \\
   24 &   233.3 &   291.7 &   350.0 &   350.0 &  1200.0 \\
   25 &   300.0 &   375.0 &   450.0 &   450.0 &  1200.0 \\
\hline
\end{tabular}
\end{center}
\end{table}

\begin{table}[h]
\small
    \caption{The discovery potential of CMS for $~L~=~10^{5}$~pb$^{-1}$
and for different signatures and for different values of 
$m_{\tilde q_3},~m_{\tilde g}$ and $tan~\beta$. All masses are in~GeV.
Here $+~(-)$ means that signal is detectable (nondetectable).
$m_{\tilde q_3}=800$,~$m_{\tilde g}=3500$,~$m_{\tilde q_{1,2}}=3800$,~ 
$\sigma$ = 0.15pb, $tan~\beta$ = 5.}
    \label{tab.3}
\begin{center}
\begin{tabular}{|c|c|c|c|c|c|c|}
\hline
 $m_{\tilde \chi_1},m_{\tilde \chi_2}$ &
 incl & no lept. & $l^{\pm} $ & $l^+l^-$ & $l^{\pm}l^{\pm}$& $3~l$ \\
\hline
133,1800 & + & + & - & - & + & - \\
\hline
400,1800 & - & - & - & - & - & - \\
\hline
600,1800 & - & - & - & - & - & - \\
\hline
720,1800 & - & - & - & - & - & - \\
\hline
 133,266 & - & - & - & - & + & - \\
\hline
 133,600 & + & + & - & - & - & - \\
\hline
 400,720 & - & - & - & - & - & - \\
\hline
 450,540 & - & - & - & - & - & - \\
\hline
\end{tabular}
\end{center}
\end{table}

\begin{table}[h]
\small
    \caption{
$m_{\tilde q_3}=750$,~$m_{\tilde g}=3500$,~$m_{\tilde q_{1,2}}=3800$,~ 
$\sigma$ = 0.2pb, $tan~\beta$ = 5.}
    \label{tab.4}
\begin{center}
\begin{tabular}{|c|c|c|c|c|c|c|}
\hline
 $m_{\tilde \chi_1},m_{\tilde \chi_2}$ &
 incl & no lept. & $l^{\pm} $ & $l^+l^-$ & $l^{\pm}l^{\pm}$& $3~l$ \\
\hline
125,1800 & + & + & - & - & + & - \\
\hline
375,1800 & - & - & - & - & + & + \\
\hline
560,1800 & - & - & - & - & - & - \\
\hline
675,1800 & - & - & - & - & - & - \\
\hline
 125,250 & - & - & - & - & + & - \\
\hline
 125,560 & + & - & - & - & + & + \\
\hline
 375,675 & - & - & - & - & - & - \\
\hline
 560,675 & - & - & - & - & - & - \\
\hline
\end{tabular}
\end{center}
\end{table}

\begin{table}[h]
\small
    \caption{
$m_{\tilde q_3}=700$,~$m_{\tilde g}=3500$,~$m_{\tilde q_{1,2}}=3800$,~ 
$\sigma$ = 0.28pb, $tan~\beta$ = 5.}
    \label{tab.5}
\begin{center}
\begin{tabular}{|c|c|c|c|c|c|c|}
\hline
 $m_{\tilde \chi_1},m_{\tilde \chi_2}$ &
 incl & no lept. & $l^{\pm} $ & $l^+l^-$ & $l^{\pm}l^{\pm}$& $3~l$ \\
\hline
117,1800 & + & + & - & - & - & - \\
\hline
350,1800 & + & + & - & - & - & - \\
\hline
525,1800 & - & - & - & - & - & - \\
\hline
630,1800 & - & - & - & - & - & - \\
\hline
 117,234 & - & - & - & + & - & - \\
\hline
 117,525 & + & + & - & - & - & - \\
\hline
 350,525 & - & - & - & - & - & - \\
\hline
 350,630 & - & - & - & - & - & - \\
\hline
 525,630 & - & - & - & - & - & - \\
\hline
\end{tabular}
\end{center}
\end{table}

\begin{table}[h]
\small
    \caption{
$m_{\tilde q_3}=650$,~$m_{\tilde g}=3500$,~$m_{\tilde q_{1,2}}=3800$,~ 
$\sigma$ = 0.42pb, $tan~\beta$ = 5.}
    \label{tab.6}
\begin{center}
\begin{tabular}{|c|c|c|c|c|c|c|}
\hline
 $m_{\tilde \chi_1},m_{\tilde \chi_2}$ &
 incl & no lept. & $l^{\pm} $ & $l^+l^-$ & $l^{\pm}l^{\pm}$& $3~l$ \\
\hline
108,1800 & + & + & - & - & - & - \\
\hline
325,1800 & + & + & - & - & - & - \\
\hline
490,1800 & - & - & - & - & - & - \\
\hline
585,1800 & - & - & - & - & - & - \\
\hline
 108,216 & + & + & - & + & + & + \\
\hline
 325,490 & - & - & - & - & - & - \\
\hline
 325,585 & + & + & - & - & - & - \\
\hline
 490,585 & - & - & - & - & - & - \\
\hline
 108,490 & + & + & - & - & - & - \\
\hline
\end{tabular}
\end{center}
\end{table}

\begin{table}[h]
\small
    \caption{
$m_{\tilde q_3}=600$,~$m_{\tilde g}=3500$,~$m_{\tilde q_{1,2}}=3800$,~ 
$\sigma$ = 0.7pb, $tan~\beta$ = 5.}
    \label{tab.7}
\begin{center}
\begin{tabular}{|c|c|c|c|c|c|c|}
\hline
 $m_{\tilde \chi_1},m_{\tilde \chi_2}$ &
 incl & no lept. & $l^{\pm} $ & $l^+l^-$ & $l^{\pm}l^{\pm}$& $3~l$ \\
\hline
100,1800 & + & + & - & - & - & - \\
\hline
300,1800 & + & + & - & - & - & - \\
\hline
450,1800 & - & - & - & - & - & - \\
\hline
540,1800 & - & - & - & - & - & - \\
\hline
 100,200 & - & - & - & - & - & - \\
\hline
 100,450 & + & + & - & - & - & - \\
\hline
 300,540 & + & + & - & - & - & - \\
\hline
 450,540 & - & - & - & - & - & - \\
\hline
 20,1800 & + & + & + & - & - & - \\
\hline
\end{tabular}
\end{center}
\end{table}

\begin{table}[h]
\small
    \caption{
$m_{\tilde q_3}=1000$,~$m_{\tilde g}=2000$,~$m_{\tilde q_{1,2}}=3800$,~ 
$\sigma$ = 0.03pb, $tan~\beta$ = 5.}
    \label{tab.8}
\begin{center}
\begin{tabular}{|c|c|c|c|c|c|c|}
\hline
 $m_{\tilde \chi_1},m_{\tilde \chi_2}$ &
 incl & no lept. & $l^{\pm} $ & $l^+l^-$ & $l^{\pm}l^{\pm}$& $3~l$ \\
\hline
170,1800 & + & + & - & - & - & - \\
\hline
500,1800 & - & - & - & - & - & - \\
\hline
750,1800 & - & - & - & - & - & - \\
\hline
900,1800 & - & - & - & - & - & - \\
\hline
 170,330 & - & - & - & - & - & - \\
\hline
 170,750 & + & + & - & - & - & - \\
\hline
 500,900 & - & - & - & - & - & - \\
\hline
 750,900 & - & - & - & - & - & - \\
\hline
\hline
\end{tabular}
\end{center}
\end{table}

\begin{table}[h]
\small
    \caption{
$m_{\tilde q_3}=1000$,~$m_{\tilde g}=1750$,~$m_{\tilde q_{1,2}}=3800$,~ 
$\sigma$ = 0.036pb, $tan~\beta$ = 5.}
    \label{tab.9}
\begin{center}
\begin{tabular}{|c|c|c|c|c|c|c|}
\hline
 $m_{\tilde \chi_1},m_{\tilde \chi_2}$ &
 incl & no lept. & $l^{\pm} $ & $l^+l^-$ & $l^{\pm}l^{\pm}$& $3~l$ \\
\hline
170,1800 & + & + & - & - & - & - \\
\hline
500,1800 & - & - & - & - & - & - \\
\hline
750,1800 & - & - & - & - & - & - \\
\hline
900,1800 & - & - & - & - & - & - \\
\hline
 170,330 & + & - & - & - & - & - \\
\hline
 170,750 & + & - & - & - & - & - \\
\hline
 500,900 & + & - & - & - & - & - \\
\hline
 750,900 & - & - & - & - & - & - \\
\hline
\hline
\end{tabular}
\end{center}
\end{table}

\begin{table}[h]
\small
    \caption{
$m_{\tilde q_3}=1000$,~$m_{\tilde g}=1500$,~$m_{\tilde q_{1,2}}=3800$,~ 
$\sigma$ = 0.038pb, $tan~\beta$ = 5.}
    \label{tab.10}
\begin{center}
\begin{tabular}{|c|c|c|c|c|c|c|}
\hline
 $m_{\tilde \chi_1},m_{\tilde \chi_2}$ &
 incl & no lept. & $l^{\pm} $ & $l^+l^-$ & $l^{\pm}l^{\pm}$& $3~l$ \\
\hline
166,1800 & + & + & - & - & - & - \\
\hline
500,1800 & + & + & - & - & - & - \\
\hline
750,1800 & - & - & - & - & - & - \\
\hline
900,1800 & - & - & - & - & - & - \\
\hline
 166,332 & + & - & + & - & - & - \\
\hline
 166,750 & + & + & + & - & + & - \\
\hline
 500,900 & + & + & - & - & - & - \\
\hline
 750,900 & - & - & - & - & - & - \\
\hline
\hline
\end{tabular}
\end{center}
\end{table}

\begin{table}[h]
\small
    \caption{
$m_{\tilde q_3}=1000$,~$m_{\tilde g}=1250$,~$m_{\tilde q_{1,2}}=3800$,~ 
$\sigma$ = 0.075pb, $tan~\beta$ = 5.}
    \label{tab.11}
\begin{center}
\begin{tabular}{|c|c|c|c|c|c|c|}
\hline
 $m_{\tilde \chi_1},m_{\tilde \chi_2}$ &
 incl & no lept. & $l^{\pm} $ & $l^+l^-$ & $l^{\pm}l^{\pm}$& $3~l$ \\
\hline
166,1800 & + & + & + & + & + & - \\
\hline
500,1800 & + & + & - & - & + & - \\
\hline
750,1800 & + & - & - & - & + & - \\
\hline
900,1800 & - & - & - & - & - & - \\
\hline
 166,332 & + & + & - & - & + & - \\
\hline
 166,750 & + & + & + & + & + & + \\
\hline
 500,900 & + & + & - & - & - & - \\
\hline
 750,900 & + & + & - & - & + & - \\
\hline
\hline
\end{tabular}
\end{center}
\end{table}

\begin{table}[h]
\small
    \caption{
$m_{\tilde q_3}=1200$,~$m_{\tilde g}=1500$,~$m_{\tilde q_{1,2}}=3800$,~ 
$\sigma$ = 0.017pb, $tan~\beta$ = 5.}
    \label{tab.12}
\begin{center}
\begin{tabular}{|c|c|c|c|c|c|c|}
\hline
 $m_{\tilde \chi_1},m_{\tilde \chi_2}$ &
 incl & no lept. & $l^{\pm} $ & $l^+l^-$ & $l^{\pm}l^{\pm}$& $3~l$ \\
\hline
200,1800 & + & + & - & - & - & - \\
\hline
600,1800 & + & - & - & - & - & - \\
\hline
900,1800 & - & - & - & - & - & - \\
\hline
1080,1800 & - & - & - & - & - & - \\
\hline
 200,400 & + & - & - & - & - & - \\
\hline
 200,600 & + & - & - & - & + & - \\
\hline
 600,900 & + & - & - & - & - & - \\
\hline
\end{tabular}
\end{center}
\end{table}

\begin{table}[h]
\small
    \caption{
$m_{\tilde q_3}=800$,~$m_{\tilde g}=2000$,~$m_{\tilde q_{1,2}}=3800$,~ 
$\sigma$ = 0.14pb, $tan~\beta$ = 5.}
    \label{tab.13}
\begin{center}
\begin{tabular}{|c|c|c|c|c|c|c|}
\hline
 $m_{\tilde \chi_1},m_{\tilde \chi_2}$ &
 incl & no lept. & $l^{\pm} $ & $l^+l^-$ & $l^{\pm}l^{\pm}$& $3~l$ \\
\hline
133,1800 & + & + & - & - & - & - \\
\hline
400,1800 & + & + & - & - & - & - \\
\hline
600,1800 & - & - & - & - & - & - \\
\hline
720,1800 & - & - & - & - & - & - \\
\hline
 133,266 & + & - & - & + & + & - \\
\hline
 133,600 & + & + & - & - & - & - \\
\hline
 600,720 & - & - & - & - & - & - \\
\hline
\end{tabular}
\end{center}
\end{table}

\begin{table}[h]
\small
    \caption{
$m_{\tilde q_3}=800$,~$m_{\tilde g}=1500$,~$m_{\tilde q_{1,2}}=3800$,~ 
$\sigma$ = 0.15pb, $tan~\beta$ = 5.}
    \label{tab.14}
\begin{center}
\begin{tabular}{|c|c|c|c|c|c|c|}
\hline
 $m_{\tilde \chi_1},m_{\tilde \chi_2}$ &
 incl & no lept. & $l^{\pm} $ & $l^+l^-$ & $l^{\pm}l^{\pm}$& $3~l$ \\
\hline
133,1800 & + & + & + & + & + & + \\
\hline
400,1800 & + & + & - & + & + & + \\
\hline
600,1800 & + & + & - & - & + & + \\
\hline
720,1800 & + & + & - & - & - & - \\
\hline
 133,266 & + & + & + & + & + & - \\
\hline
 133,600 & + & + & + & + & + & + \\
\hline
\end{tabular}
\end{center}
\end{table}

\begin{table}[h]
\small
    \caption{
$m_{\tilde q_3}=800$,~$m_{\tilde g}=1000$,~$m_{\tilde q_{1,2}}=3800$,~ 
$tan~\beta$ = 5.}
    \label{tab.15}
\begin{center}
\begin{tabular}{|c|c|c|c|c|c|c|}
\hline
 $m_{\tilde \chi_1},m_{\tilde \chi_2}$ &
 incl & no lept. & $l^{\pm} $ & $l^+l^-$ & $l^{\pm}l^{\pm}$& $3~l$ \\
\hline
133,1800 & + & + & + & + & + & + \\
\hline
400,1800 & + & + & + & - & + & + \\
\hline
600,1800 & + & + & - & - & + & - \\
\hline
720,1800 & + & + & - & - & + & - \\
\hline
 133,266 & + & + & + & + & + & + \\
\hline
 133,600 & + & + & + & + & + & + \\
\hline
 400,720 & + & + & - & - & + & + \\
\hline
 600,720 & + & + & - & - & + & - \\
\hline
\end{tabular}
\end{center}
\end{table}

\begin{table}[h]
\small
    \caption{
$m_{\tilde q_3}=3800$,~$m_{\tilde g}=3500$,~$m_{o3}=1000$,~ 
$\sigma$ = 0.03pb, $tan~\beta$ = 35.}
    \label{tab.16}
\begin{center}
\begin{tabular}{|c|c|c|c|c|c|c|}
\hline
 $m_{\tilde \chi_1},m_{\tilde \chi_2}$ &
 incl & no lept. & $l^{\pm} $ & $l^+l^-$ & $l^{\pm}l^{\pm}$& $3~l$ \\
\hline
166,1800 & + & - & - & - & - & - \\
\hline
500,1800 & - & - & - & - & - & - \\
\hline
750,1800 & - & - & - & - & - & - \\
\hline
850,1800 & - & - & - & - & - & - \\
\hline
 166,322 & - & - & - & - & - & - \\
\hline
 166,750 & + & - & - & - & - & - \\
\hline
 500,750 & - & - & - & - & - & - \\
\hline
 500,900 & - & - & - & - & - & - \\
\hline
 750,900 & - & - & - & - & - & - \\
\hline
\end{tabular}
\end{center}
\end{table}

\clearpage

\begin{table}[h]
\small
    \caption{
$m_{\tilde q_3}=3800$,~$m_{\tilde g}=3500$,~$m_{o3}=900$,~ 
$\sigma$ = 0.07pb, $tan~\beta$ = 35.}
    \label{tab.17}
\begin{center}
\begin{tabular}{|c|c|c|c|c|c|c|}
\hline
 $m_{\tilde \chi_1},m_{\tilde \chi_2}$ &
 incl & no lept. & $l^{\pm} $ & $l^+l^-$ & $l^{\pm}l^{\pm}$& $3~l$ \\
\hline
150,1800 & + & + & - & - & - & - \\
\hline
450,1800 & - & - & - & - & - & - \\
\hline
675,1800 & - & - & - & - & - & - \\
\hline
750,1800 & - & - & - & - & - & - \\
\hline
 150,300 & - & - & - & - & + & - \\
\hline
 150,675 & + & + & - & - & - & - \\
\hline
 450,675 & - & - & - & - & - & - \\
\hline
 450,810 & - & - & - & - & - & - \\
\hline
 675,810 & - & - & - & - & - & - \\
\hline
\end{tabular}
\end{center}
\end{table}

\begin{table}[h]
\small
    \caption{
$m_{\tilde q_{1,2}}=3800$,~$m_{\tilde g}=3500$,~$m_{o3}=800$,~ 
$\sigma$ = 0.18pb, $tan~\beta$ = 35.}
    \label{tab.18}
\begin{center}
\begin{tabular}{|c|c|c|c|c|c|c|}
\hline
 $m_{\tilde \chi_1},m_{\tilde \chi_2}$ &
 incl & no lept. & $l^{\pm} $ & $l^+l^-$ & $l^{\pm}l^{\pm}$& $3~l$ \\
\hline
133,1800 & + & + & - & - & - & - \\
\hline
400,1800 & - & - & - & - & - & - \\
\hline
600,1800 & - & - & - & - & - & - \\
\hline
720,1800 & + & + & + & - & - & - \\
\hline
 133,266 & - & - & - & - & + & + \\
\hline
 133,600 & + & + & - & + & + & + \\
\hline
 400,600 & - & - & - & - & - & - \\
\hline
 400,720 & - & - & - & - & - & - \\
\hline
 450,540 & - & - & - & - & + & + \\
\hline
\end{tabular}
\end{center}
\end{table}

\begin{table}[h]
\small
    \caption{
$m_{\tilde q_3}=750$,~$m_{\tilde g}=3500$,~$m_{\tilde q_{1,2}}=3800$,~ 
$\sigma$ = 0.28pb, $tan~\beta$ = 35.}
    \label{tab.19}
\begin{center}
\begin{tabular}{|c|c|c|c|c|c|c|}
\hline
 $m_{\tilde \chi_1},m_{\tilde \chi_2}$ &
 incl & no lept. & $l^{\pm} $ & $l^+l^-$ & $l^{\pm}l^{\pm}$& $3~l$ \\
\hline
125,1800 & + & + & + & + & - & - \\
\hline
375,1800 & + & + & + & + & - & - \\
\hline
560,1800 & + & + & + & + & - & - \\
\hline
675,1800 & + & + & + & + & - & - \\
\hline
 125,250 & + & + & + & + & + & + \\
\hline
 125,560 & + & + & + & + & + & + \\
\hline
 375,560 & + & + & - & - & + & - \\
\hline
 560,675 & - & - & - & - & - & - \\
\hline
\end{tabular}
\end{center}
\end{table}

\begin{table}[h]
\small
    \caption{
$m_{\tilde q_3}=650$,~$m_{\tilde g}=3500$,~$m_{\tilde q_{1,2}}=3800$,~ 
$\sigma$ = 1pb, $tan~\beta$ = 35.}
    \label{tab.20}
\begin{center}
\begin{tabular}{|c|c|c|c|c|c|c|}
\hline
 $m_{\tilde \chi_1},m_{\tilde \chi_2}$ &
 incl & no lept. & $l^{\pm} $ & $l^+l^-$ & $l^{\pm}l^{\pm}$& $3~l$ \\
\hline
108,1800 & + & + & - & + & - & - \\
\hline
325,1800 & + & + & + & + & - & - \\
\hline
487,1800 & + & + & + & + & - & - \\
\hline
585,1800 & + & + & + & + & - & - \\
\hline
 108,216 & - & - & - & + & + & + \\
\hline
 108,487 & + & + & - & + & - & + \\
\hline
 487,585 & + & + & + & + & - & + \\
\hline
\end{tabular}
\end{center}
\end{table}

\begin{table}[h]
\small
    \caption{
$m_{o3}=1200$,~$m_{\tilde g}=1500$,~$m_{\tilde q_{1,2}}=3800$,~ 
$\sigma$ = 0.02pb, $tan~\beta$ = 35.}
    \label{tab.21}
\begin{center}
\begin{tabular}{|c|c|c|c|c|c|c|}
\hline
 $m_{\tilde \chi_1},m_{\tilde \chi_2}$ &
 incl & no lept. & $l^{\pm} $ & $l^+l^-$ & $l^{\pm}l^{\pm}$& $3~l$ \\
\hline
200,1800 & + & + & - & - & - & - \\
\hline
600,1800 & + & - & - & - & - & - \\
\hline
900,1800 & - & - & - & - & - & - \\
\hline
1080,1800 & + & + & - & - & - & - \\
\hline
 200,400 & + & - & - & - & - & - \\
\hline
 200,600 & + & - & + & - & + & - \\
\hline
 600,900 & + & - & - & - & - & - \\
\hline
 900,1080 & - & - & - & - & - & - \\
\hline
\end{tabular}
\end{center}
\end{table}

\begin{table}[h]
\small
    \caption{
$m_{o3}=800$,~$m_{\tilde g}=1000$,~$m_{\tilde q_{1,2}}=3800$,~ 
$\sigma$ = 0.5pb, $tan~\beta$ = 35.}
    \label{tab.22}
\begin{center}
\begin{tabular}{|c|c|c|c|c|c|c|}
\hline
 $m_{\tilde \chi_1},m_{\tilde \chi_2}$ &
 incl & no lept. & $l^{\pm} $ & $l^+l^-$ & $l^{\pm}l^{\pm}$& $3~l$ \\
\hline
133,1800 & + & + & + & + & + & + \\
\hline
400,1800 & + & + & + & + & + & - \\
\hline
600,1800 & + & + & - & - & + & - \\
\hline
720,1800 & + & + & + & + & + & + \\
\hline
 133,266 & + & + & + & + & + & + \\
\hline
 133,600 & + & + & + & + & + & + \\
\hline
 400,600 & + & + & + & + & + & + \\
\hline
 400,720 & + & + & + & + & + & - \\
\hline
\end{tabular}
\end{center}
\end{table}

\begin{table}[h]
\small
    \caption{
$m_{o3}=800$,~$m_{\tilde g}=1500$,~$m_{\tilde q_{1,2}}=3800$,~ 
$tan~\beta$ = 35.}
    \label{tab.23}
\begin{center}
\begin{tabular}{|c|c|c|c|c|c|c|}
\hline
 $m_{\tilde \chi_1},m_{\tilde \chi_2}$ &
 incl & no lept. & $l^{\pm} $ & $l^+l^-$ & $l^{\pm}l^{\pm}$& $3~l$ \\
\hline
133,1800 & + & + & - & - & + & - \\
\hline
400,1800 & + & + & - & - & - & - \\
\hline
600,1800 & - & - & - & - & + & - \\
\hline
720,1800 & + & + & - & - & - & - \\
\hline
 133,266 & + & + & + & + & + & + \\
\hline
 133,600 & + & + & + & + & + & + \\
\hline
 400,600 & + & + & - & - & + & - \\
\hline
 400,720 & + & - & - & + & + & + \\
\hline
 600,720 & + & - & - & + & + & + \\
\hline
\end{tabular}
\end{center}
\end{table}

\begin{table}[h]
\small
    \caption{
$m_{\tilde g}=3500$,~$m_{\tilde q_{1,2}}=3800$.
}
    \label{tab.24}
\begin{center}
\begin{tabular}{|c|c|c|c|c|c|c|c|c|}
\hline
 $m_{\tilde \chi_1},m_{\tilde \chi_2}$ & $tan~\beta$ &  & 
 incl & no lept. & $l^{\pm} $ & $l^+l^-$ & $l^{\pm}l^{\pm}$& $3~l$ \\
\hline
166,1800 & 5 & $m_{\tilde q_3}$=1000 & + & + & - & - & - & - \\
\hline
$\frac{m_{\tilde q_3}}{6}$,1800 &
              5 & $m_{\tilde q_3}$=1100 & - & - & - & - & - & - \\
\hline
$\frac{m_{\tilde q_3}}{6}$,1800 &
              5 & $m_{\tilde q_3}$=1200 & - & - & - & - & - & - \\
\hline
$\frac{m_{\tilde q_3}}{6}$,1800 &
              5 & $m_{\tilde q_3}$=1300 & - & - & - & - & - & - \\
\hline
$\frac{m_{0}}{6}$,1800 &
              35 & $m_{0}$=1000 & - & - & - & - & - & - \\
\hline
$\frac{m_{0}}{6}$,1800 &
              35 & $m_{0}$=1100 & - & - & - & - & - & - \\
\hline
$\frac{m_{0}}{6}$,1800 &
              35 & $m_{0}$=1200 & - & - & - & - & - & - \\
\hline
$\frac{m_{0}}{6}$,1800 &
              35 & $m_{0}$=1300 & - & - & - & - & - & - \\
\hline
\end{tabular}
\end{center}
\end{table}

\end{document}